\documentclass[pra,a4paper,twoside,superscriptaddress,11pt,tightenlines]{revtex4}
\usepackage{times,amssymb,graphicx,bbm,amsmath}
\usepackage[dvips]{color}
\begin{document}
\title{Reconstruction of photon statistics using 
low performance photon counters}
\author{Guido Zambra}
\affiliation{Dipartimento di Fisica dell'Universit\`a degli Studi di Milano, Italia.}
\affiliation{Dipartimento di Fisica e Matematica dell'Universit\`a
degli Studi dell'Insubria, Como, Italia.}
\author{Matteo G. A. Paris}
\affiliation{Dipartimento di Fisica dell'Universit\`a degli Studi di Milano, Italia.}
\begin{abstract}
The output of a photodetector consists of a current pulse whose charge has the
statistical distribution of the actual photon numbers convolved with a
Bernoulli distribution.  Photodetectors are characterized by a nonunit quantum
efficiency, {\em i.e.} not all the photons lead to a charge, and by a finite
resolution, {\em i.e.} a different number of detected photons leads to a
discriminable values of the charge only up to a maximum value.  We present a
detailed comparison, based on Monte Carlo simulated experiments and real data,
among the performances of detectors with different upper limits of counting
capability.  In our scheme the inversion of Bernoulli convolution is performed
by maximum-likelihood methods assisted by measurements taken at different
quantum efficiencies. We show that detectors that are only able to
discriminate between zero, one and more than one detected photons are
generally enough to provide a reliable reconstruction of the photon statistics
for single-peaked distributions, while detectors with higher resolution limits
do not lead to further improvements. In addition, we demonstrate that, for
semiclassical states, even on/off detectors are enough to provide a good
reconstruction. Finally, we show that a reliable reconstruction of
multi-peaked distributions requires either higher quantum efficiency or better
capability in discriminating high number of detected photons.  
\end{abstract}
\date{\today}
\maketitle
\section{Introduction}
Reconstruction of the photon statistics, ${\varrho_n}$, of optical
states provides fundamental information on the nature of any
optical field and finds relevant applications in foundations of
quantum mechanics, quantum state engineering by
postselection \cite{pst}, quantum information \cite{post}, and quantum
metrology.  Indeed, detectors with the capability of counting
photons \cite{rivelacontaS,rivelacontaH} are currently under
investigation.  Among these, photomultiplier tubes (PMT's) 
\cite{burle} and hybrid photodetectors \cite{rivelacontaH,NIST} 
are promising devices, though they have the drawback of a low 
quantum efficiency. On the other hand, solid state
detectors with internal gain are still under development.  
Highly efficient thermal detectors have also been used
as photon counters, though their operating conditions are
still extreme to allow common use \cite{xxx,serg}.
The advent of quantum tomography provided an alternative
method to measure photon number distributions \cite{mun}.
However, tomography needs the implementation of homodyne
detection, which involves challenging mode matching, 
especially in the case of pulsed optical fields.
\par
In principle, in a photodetector each photon ionizes a single
atom, and the resulting charge is amplified to produce a
measurable pulse. In practice, however, available photodetectors
are usually characterized by a quantum efficiency lower than
unity, which means that only a fraction of the incoming photons
leads to an electric pulse. If the resulting current is proportional to the
incoming photon flux we have a linear photodetector. This is,
for example, the case of the high-flux photodetectors used in
homodyne detection.  On the other hand photodetectors operating
at very low intensities usually resort to avalanche process or very
high amplification in order to transform a single ionization event
into a recordable pulse. Due to the gain instability in the process
of amplification it is generally difficult to discriminate the
number of detected photons as far as this number becomes larger.
\par 
As a consequence the output of a photodetector consists of a current
pulses whose charge statistics is a generalized Bernoulli
convolution of the actual photon distribution with the possibility to
perform a discrimination between the number of detected photons
only up to a finite number.
We define $M$ as the maximum number $m$ of detected photons that can be
distinguished from $m-1$, $M$ represent the counting capability of the
detectors: ${\emph i.e.}$ detectors that works in Geiger mode have
$M=1$ because form the output is possible to discriminate from $m=1$, one
detected photon, to $m=0$, dark, but is not possible to discriminate
the output for $m\geq2$ from that for to $m=1$.
The inversion of such a convolution may be performed in several
ways, though, in general it is possible only when the quantum
efficiency is larger than $\eta=0.5$ \cite{deconvolvphoton}, and it is 
inherently inefficient, as it requires a large data sample to 
provide a reliable result.  On the other hand, maximum-likelihood
methods assisted by measurements taken at different quantum
efficiencies have been proved to be both effective and
statistically reliable \cite{mogy,pcount,cvp,olom}.  
In particular, it is
possible to obtain a method  to reconstruct ${\varrho_{n}}$ {\em
without any a priori information} on the state of light under
investigation.  An alternative approach is based on the so called
{\em photon chopping} \cite{phc}, {\em i.e.} on a network of beam splitter
followed by an array of on/off detectors. Photon chopping allows 
to send at most one photon on each detector, however, with the 
drawback of increasing the overall complexity of the detection 
scheme.
\par
In this paper, we extend previous analysis on maximum-likelihood
on/off-based schemes \cite{pcount,cvp}, 
and present a detailed comparison among the
performances of detectors characterized by realistic quantum
efficiencies and different values of $M$. The comparison
is made using real data and by an extensive set of Monte Carlo
simulations, performed on different states of the radiation field
including quantum and semiclassical states associated to
single-peaked as well as multi-peaked photon distributions. The
reconstruction is obtained by maximum-likelihood methods assisted
by measurements taken at different quantum efficiencies. In
particular, as the statistics of the detected photons is linearly dependent
on the photon distribution (see below), the inversion can be well
approximated by using the Expectation-Maximization algorithm (EM)
which leads to an effective and reliable iterative solution.
\par
Our results indicates that detectors with $M=2$ discriminating
between zero, one and more than one photon, are generally enough
to provide a reliable reconstruction of the photon statistics for
single-peaked distributions, while detectors with higher value of $M$
do not lead to further improvements. On the other hand,
multi-peaked distributions requires a higher quantum efficiency
whereas for semiclassical states even on/off detectors are enough
to provide a good reconstruction. 
\par
The paper is structured as follows. In the next Section 
we describe the statistics of the detected photons and illustrate
the reconstruction algorithm.
In Section \ref{s:sim} we report the results of a set
of Monte Carlo simulated experiments performed on different
kinds of signals, whereas in Section \ref{s:exp} we report
experimental results for coherent signals. 
Section \ref{s:out} closes the paper with some concluding remarks.
\section{Statistics of detected photons and reconstruction algorithm}
Using a photodetector with quantum efficiency $\eta$ and $M=\infty$,
that corresponding to an unlimited counting capability,
the probability of obtaining $m$ detected photons at the output 
is given by the convolution 
\begin{equation}\label{convoluzio0}
p_\eta(m)=\sum^\infty_{n=m} A_\nu(m,n) \varrho_n\:,
\end{equation}
where $\varrho_n= \langle n |\varrho | n\rangle$ is the
actual photon statistics of the signal under investigation 
and 
\begin{equation}
A_\eta (m,n) = \left(\begin{array}{c}n\\m\end{array}\right)(1-\eta)^{n-m}\eta^m  \:.
\end{equation}
If $M$ have a finite value only $M+1$ outcomes are possible, which occur
with probabilities 
\begin{align}
q_\eta^m &= p_\eta(m) \quad m=0,...,M-1  \label{kappa1}\\
q_\eta^M &= \sum_{m=M}^\infty p_\eta(m)  = 1 - \sum_{m=0}^{M-1} q_\eta^m
\label{kappa2}\;.
\end{align}
Once the value of the quantum efficiency is known, Eqs. 
(\ref{kappa1}) and (\ref{kappa2}) provide $M$ relations
among the statistics of detected photons and the actual statistics
of photons. 
At first sight the statistics of a detector with low counting
capability ($M$ in the range $M=1,...,6$)
appears to provide quite a scarce piece of information about the
state under investigation. However, if the statistics about
$\{q_\eta^m \}$ is collected for a suitably large set of efficiency
values, then the information is enough to reconstruct the whole
photon distribution $\varrho_n$ of the signal, upon a suitable
truncation at $N$ of the Hilbert space.
We adopt the following strategy: by
placing in front of the detector $K$ filters with different
transmissions, we may perform the detection with $K$ different 
values $\eta_\nu$, $\nu=1...K$, ranging from zero to a maximum
value $\eta_K=\eta_{\max}$
equal to the nominal quantum efficiency of the detector. 
By denoting the probability of having $m$ detected photons
in the experiment with quantum efficiency $\eta=\eta_\nu$ 
by $q_\nu^m\equiv q_{\eta_\nu}^m$
we can rewrite Eqs. (\ref{kappa1}) and (\ref{convoluzio0}) as 
\begin{equation}\label{convoluzio}
q_\nu^m= \sum^\infty_{n=m} A_\nu (m,n) \varrho_n 
\:,
\end{equation}
where $\nu=1,..K$ and $m=0,..,M-1$. 
Let us now suppose that the $\varrho_n$'s are negligible for 
$n>N$ and that the $\eta_\nu$'s are known, then $\forall \eta_\nu$ 
Eq.~(\ref{convoluzio}) may be rewritten a finite sum over $n$ 
from $m$ to $N$. Overall, we obtain a finite linear system with 
$K\times M$ equations in the $N$ unknowns $\{\varrho_n\}$.  
Unfortunately, the
reconstruction of $\varrho_n$ by matrix inversion cannot be used in
practice since it would require an unreasonable number of experimental runs
to assure the necessary precision \cite{mogy,pcount}. This problem can be 
circumvented by considering
Eqs.~(\ref{convoluzio}) as a statistical model for the parameters
$\{\varrho_n\}$ to be solved by maximum-likelihood (ML) estimation. 
The likelihood functional is given by the global probability of the 
sample {\em i.e}
\begin{eqnarray}
{\sf L} = \prod_{\nu=1}^K \prod_{m=0}^{M-1} \left( q_\nu^m \right)^{n_{m\nu}}
\label{Lik}\;.
\end{eqnarray}
The ML estimates of $\{\varrho_n\}$ are the values that maximizes 
${\sf L}$.
In Eq. (\ref{Lik}) $n_{m\nu}$ denotes the number of events "$m$ detected
photons" obtained with quantum efficiency $\eta_\nu$.
The maximization of ${\sf L}$ under the conditions $\varrho_n\geq 0$, 
$\sum_n\varrho_n=1$, can be well approximated by
using the expectation-maximization (EM) algorithm \cite{EM:alg:1,
EM:alg:2,kon}, which leads to the iterative solution
\begin{align}\label{iterazio}
\varrho^{(i+1)}_n =&\varrho^{(i)}_n \left(\sum_{\nu=1}^K\sum_{m=0}^M
A_\nu(m,n)\right)^{-1}
\nonumber \\ &\times
\sum_{\nu=1}^K\sum_{m=0}^M
A_\nu(m,n)\:\frac{f_\nu^m}{q_\nu^m[\{\varrho^{(i)}_n\}]}\:.
\end{align}
In Eq. (\ref{iterazio}) $\varrho_n^{(i)}$ denotes the $n$-th
element of reconstructed 
statistics at the $i$-th step, $q_\nu^m[\{\varrho^{(i)}_n\}]$ 
the theoretical probabilities as calculated from Eq. (\ref{convoluzio})
at the $i$-th step, whereas 
$f_\nu^m= n_{m\nu}/n_\nu$ represents the frequency
of the event "$m$ detected photons" with quantum efficiency
$\eta_\nu$, being $n_\nu$
the total number of runs performed with $\eta=\eta_\nu$. In the following
we assume equal $n_\nu$, $\forall \nu$.
\par
Eq. (\ref{iterazio}) provides a solution once an initial 
distribution is chosen. In the following we will always consider
an initial uniform distribution $\varrho_n^{(0)}=(1+N)^{-1}$ $\forall n$ in 
the truncated Fock space $n=0,...,N$. Results from Monte Carlo
simulated experiments shows that any distribution with 
$\varrho_n^{(0)}\neq0$ $\forall n$ performs equally well, {\em i.e.} 
the choice of the initial distribution does not alter
the quality of reconstruction, though it may slightly affect 
the convergence properties of the algorithm.
\section{Monte Carlo simulated experiments}\label{s:sim}
In this Section we report the photon statistics for different 
states of a single-mode radiation field as obtained by ML 
reconstruction from Monte Carlo simulated photodetection 
performed with the same low quantum efficiency and different resolution
threshold.  We consider semiclassical states as well as 
highly nonclassical states 
and show results for different values of the maximal 
quantum efficiency $\eta_{\max}$.
\begin{widetext} $ $ 
\begin{figure}[h]
\centering
\begin{tabular}{ccc}
\includegraphics[width=0.22\textwidth]{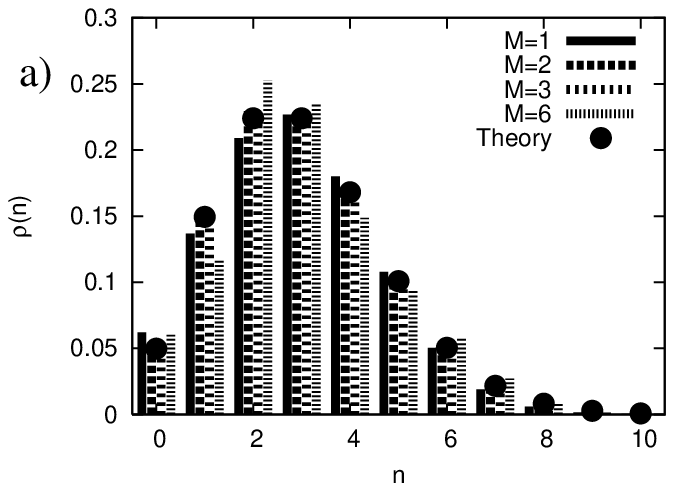}&
\includegraphics[width=0.22\textwidth]{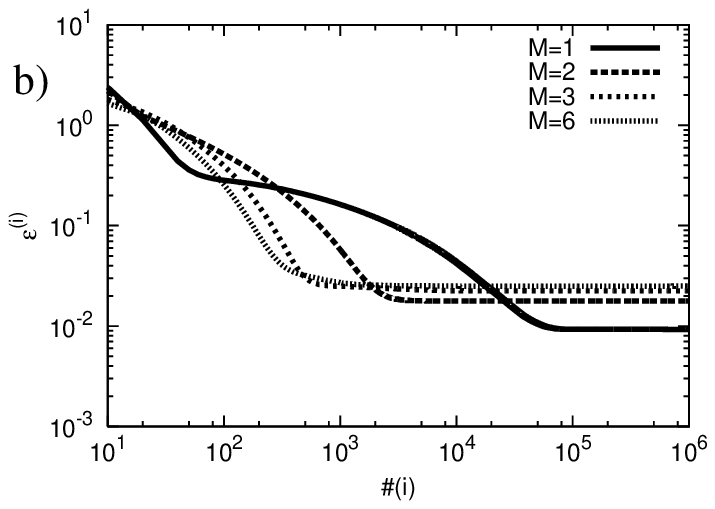}&
\includegraphics[width=0.22\textwidth]{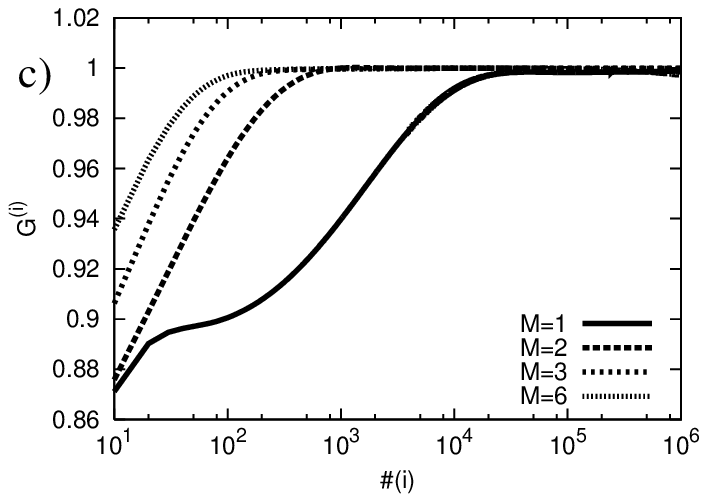}\\
\includegraphics[width=0.22\textwidth]{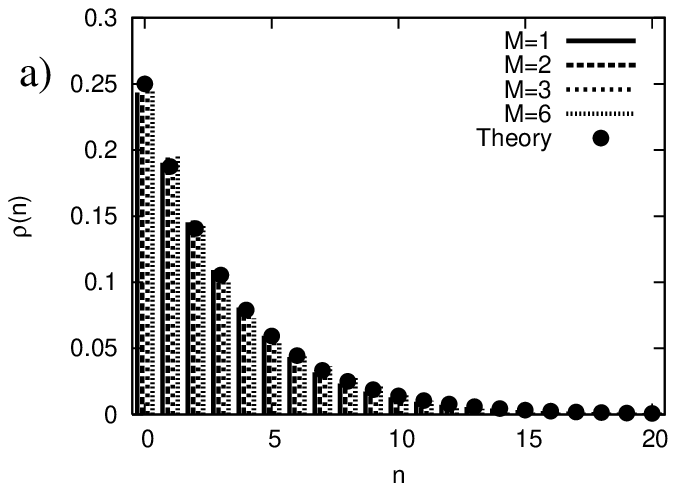}&
\includegraphics[width=0.22\textwidth]{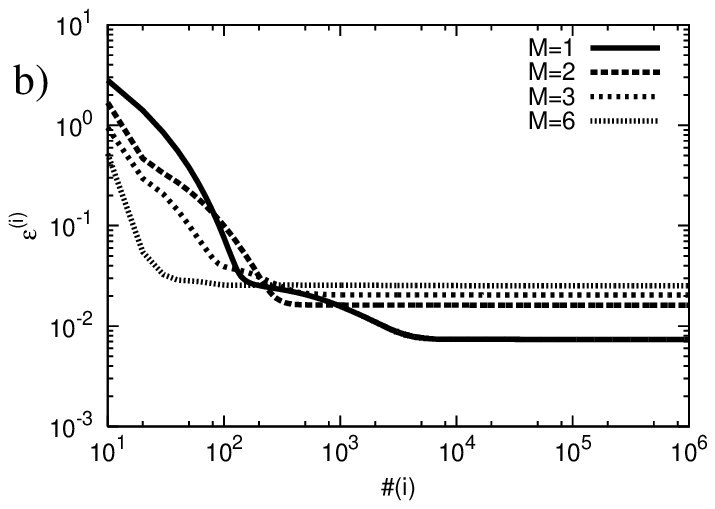}&
\includegraphics[width=0.22\textwidth]{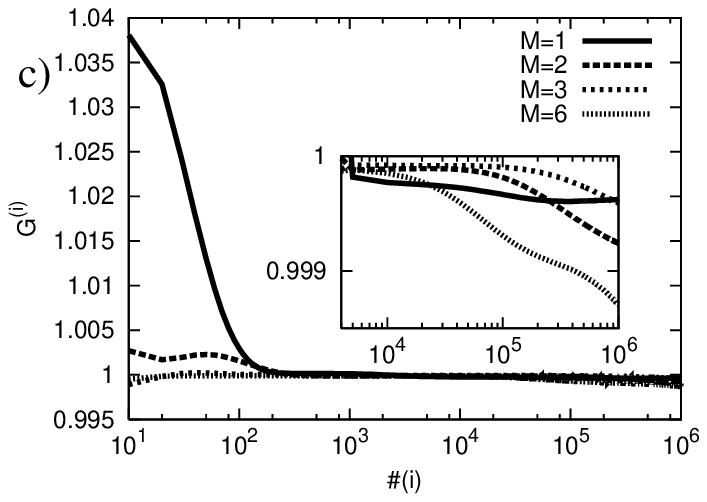}\\
\includegraphics[width=0.22\textwidth]{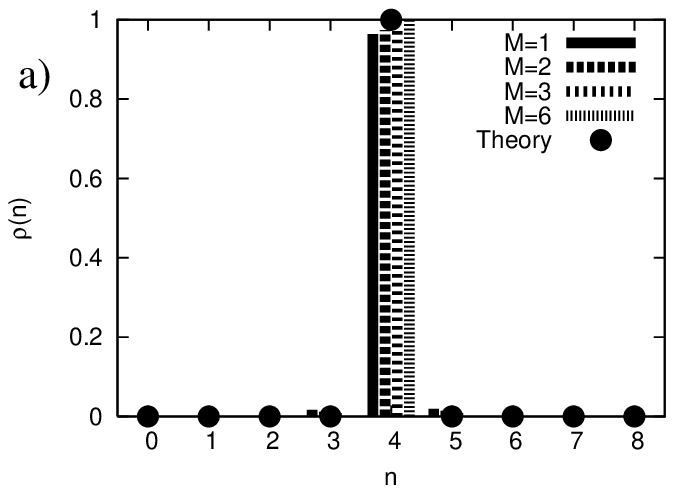}&
\includegraphics[width=0.22\textwidth]{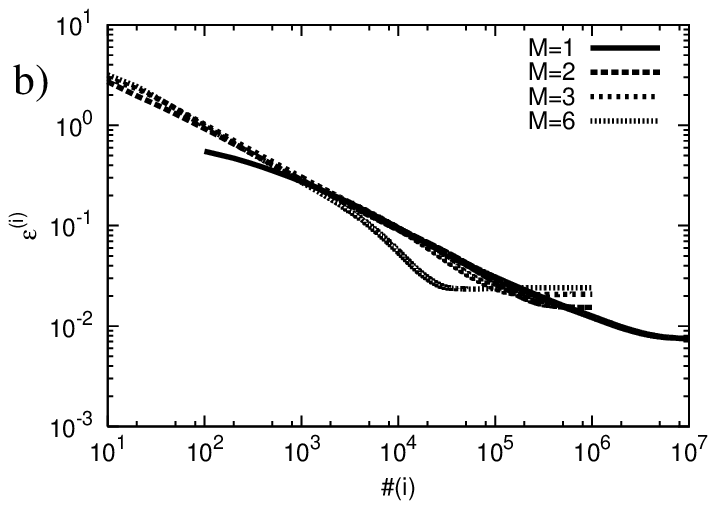}&
\includegraphics[width=0.22\textwidth]{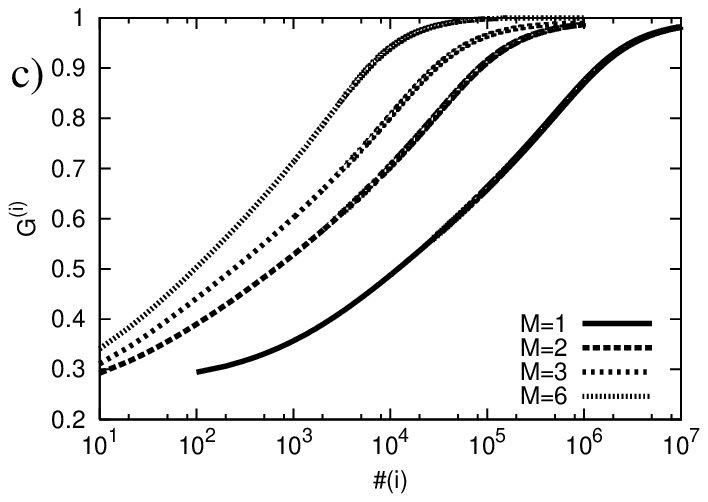}\\
\includegraphics[width=0.22\textwidth]{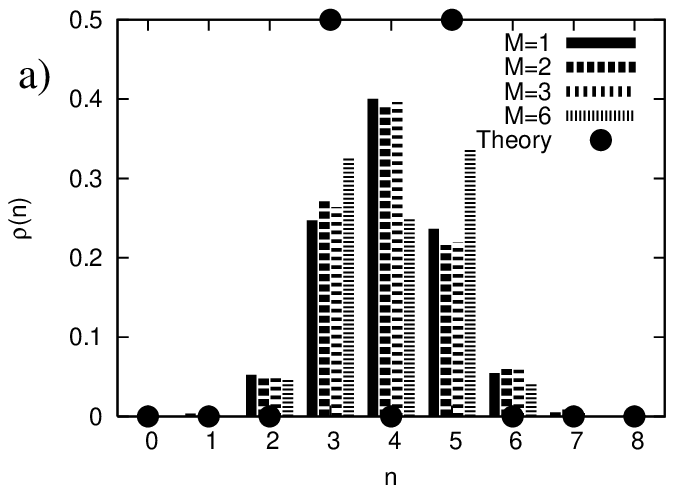}&
\includegraphics[width=0.22\textwidth]{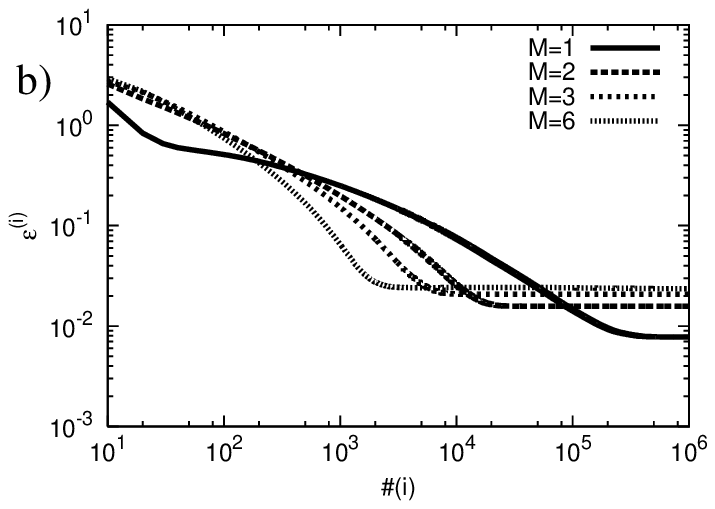}&
\includegraphics[width=0.22\textwidth]{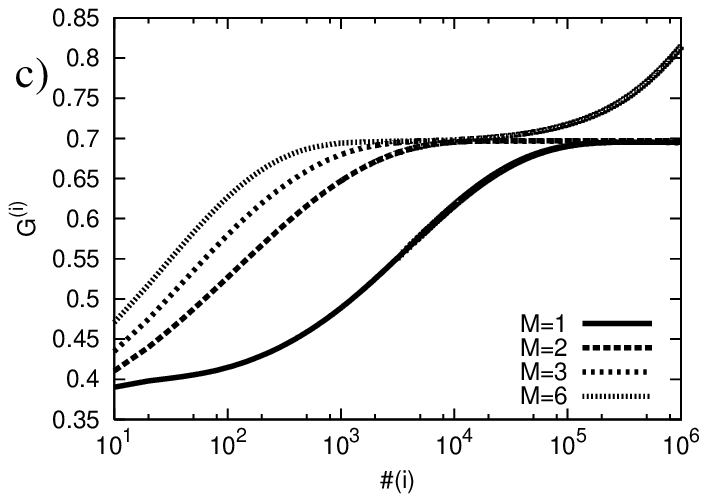}
\end{tabular}
\caption{Maximum-likelihood reconstruction of photon statistics using 
photodetectors with different counting capability (different values of $M$)
and maximum quantum efficiency $\eta_{max}=0.2$. 
First line: photon distribution of 
a coherent state with $\langle a^\dag a\rangle=3$; panel a) 
the reconstructed $\rho_n$; b): total absolute error $\varepsilon^{(i)}$; 
according to Eq. (\ref{epsilon})
c): fidelity $G^{(i)}$ according to Eq. (\ref{fidelity}). Results 
are reported for $M=1$ (on/off detector), $M=2$, $M=3$ and $M=6$. 
We use $K=30$ different quantum efficiencies 
$\eta=\eta_\nu$, uniformly distributed in $[\eta_{max}/K,\eta_{max}]$.
The Hilbert space have been truncated at $N=30$ 
and $n_\nu=10^6$ runs have been performed for each value 
$\eta_\nu$. The last iteration is $\#(i_{L})=n_\nu$.
Other lines: same as first line for thermal 
state with $\langle a^\dag a\rangle=3$, Fock state with $n=4$;
superposition of 2 Fock states $\frac{1}{\sqrt{2}}(|3\rangle+|5\rangle)$.
The inset of panel c) in the third line shows the last iterations 
$\#(i)$ in an expanded scale.}\label{fig:pois1e6}
\end{figure}
\end{widetext}
In order to assess our results we introduce two figures 
of merit, to measure  the reliability and the accuracy 
of the method respectively. Since the solution of the ML 
estimation is obtained iteratively,
the most important aspect to keep under control is the
convergence of the algorithm. A suitable parameter to
evaluate the degree of convergence at the $i$-th iteration
is the total absolute error
\begin{equation} \label{epsilon}
\varepsilon^{(i)} =\sum_{\nu=1}^K \sum_{m=0}^{M}  \left|
f_\nu^m - q_\nu^m[\{\varrho^{(i)}_n\}]
\right|\:.
\end{equation}
Indeed,  the total error measures the distance of the
probabilities $q_\nu^m [\{ \varrho_n^{(i)}\}]$, as calculated at
the $i$-th iteration, from the actual experimental frequencies
and thus, besides convergence, it quantifies how the estimated
distribution reproduces the experimental data. The total distance
is a decreasing function of the number of iterations. Its
stationary value is proportional to the accuracy of the
experimental frequencies $\{f_\nu^m\}$.  For finite data sample
this value is of order $1/\sqrt{n_\nu}$ for each value of
$\eta_\nu$.  As a measure of accuracy at the $i$-th step we adopt
the so-called fidelity  between probability distributions
\begin{eqnarray} \label{fidelity}
G^{(i)} = \sum_{n=0}^{N-1}
\sqrt{\varrho^{(i)}_n\varrho_n}\:. 
\end{eqnarray} 
In Ref. \cite{cvp} it was shown that using on/off detection ({\em
i.e.} $M=1$) a reliable reconstruction
scheme may be obtained. In this paper, our aim is to check
whether a higher value of $M$ leads to some advantages,
either in terms of accuracy or convergence.  
\par 
Fig. \ref{fig:pois1e6} summarizes results
for different states and values of $M$, assuming
$\eta_{max}=0.2$ as the maximum value of the quantum efficiency.
In order to compare the performances achievable by different
$M$ one has first to chose a criterion to 
stop the iterative algorithm. A natural criterion would be that 
of stopping the iteration when the total absolute error 
$\varepsilon^{(i)}$ converges, {\em i.e} when the rate of its 
variation falls below a threshold and  becomes negligible.  
This is also motivated by the behavior of $\varepsilon^{(i)}$ 
versus the number of iterations (see central column in Fig. 
\ref{fig:pois1e6}): a rapid fall followed by a plateau.
However, the most convenient value for the threshold 
unavoidably depends on the shape of the unknown distribution 
$\varrho_{(n)}$. Therefore, in the absence of any \textit{a priori} 
information, this simple criterion should be supplemented by 
some additional recipes.  We found numerically that a suitable 
criterion, valid for any value of $M$ and, in the average, for 
any class of states, is the following: if the total absolute 
error appears to converge, then stop the algorithm at a number 
of iterations $\#(i)=n_\nu$ equal to the number of measurements 
taken at each value of the quantum efficiency, otherwise stop the 
algorithm as soon as you see convergence.
In cases when $\varepsilon^{(i)}$ converges for a number of iteration 
$\#(i) < n_\nu$ it is not convenient to stop the algorithm, since
it may further increase the quality of the reconstruction, see for 
example $G^{(i)}$ in the fourth line of Fig \ref{fig:pois1e6}.
On other hands, if $\#(i)$ goes far beyond the condition of
convergence of $\varepsilon^{(i)}$, then the  algorithm may lose 
precision due to the noise of the data $f_\nu^m$ (see inset in 
the second line of Fig. \ref{fig:pois1e6}). 
The choice of the threshold at $\#(i)=n_\nu$ fits with other
two independent facts. On one hand, we performed an extensive 
set of simulations and no considerable reduction of $G^{(i)}$ 
was observed before the convergence condition $\#(i)=n_\nu$. 
On the other hand, we applied the reconstruction algorithm to 
the {\em exact} on/off frequencies (not to simulated data) 
and no reduction of $G^{(i)}$ was observed for increasing 
$\#(i)$, thus confirming the data-noise origin of precision loss. 
\begin{widetext}  $ $ 
\begin{figure}[h!]
\centering
\begin{tabular}{ccc}
\includegraphics[width=0.22\textwidth]{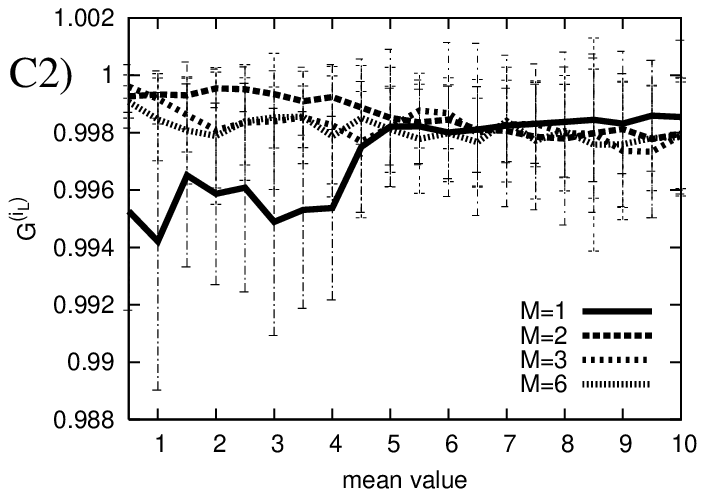}&
\includegraphics[width=0.22\textwidth]{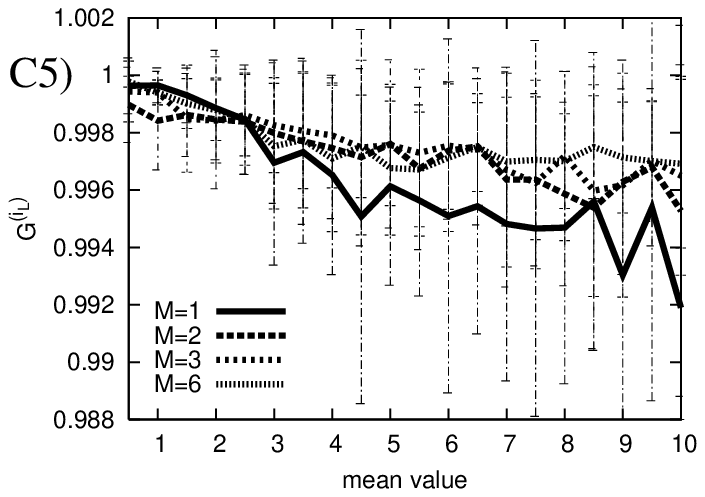}&
\includegraphics[width=0.22\textwidth]{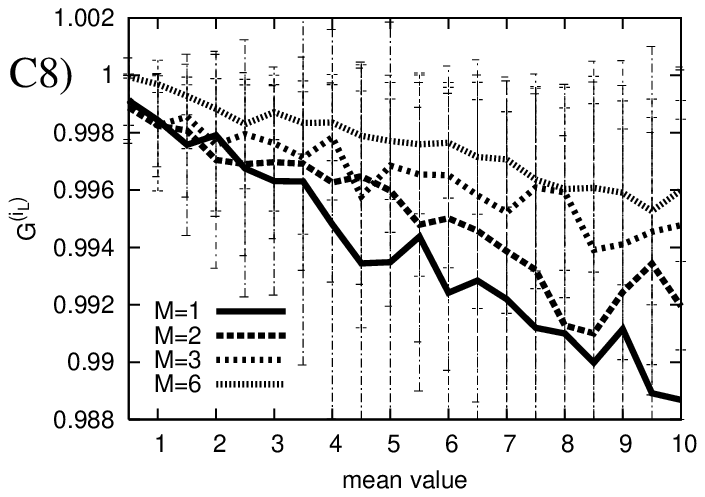}\\
\includegraphics[width=0.22\textwidth]{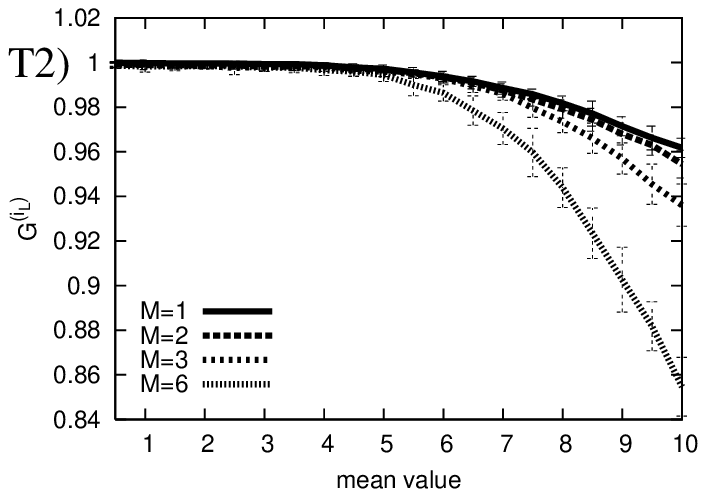}&
\includegraphics[width=0.22\textwidth]{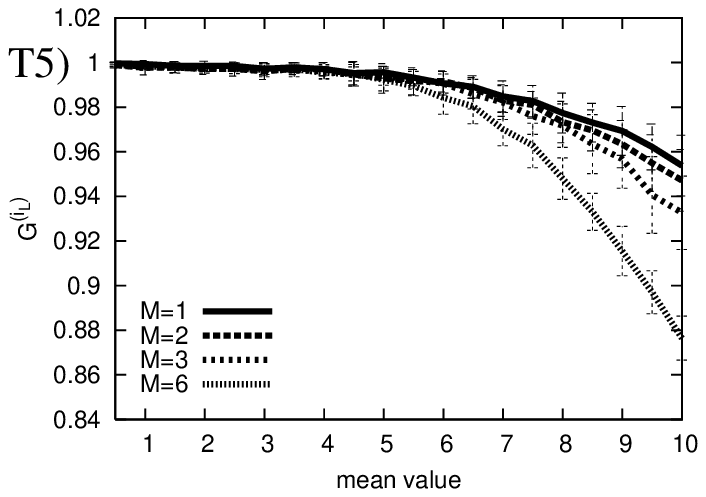}&
\includegraphics[width=0.22\textwidth]{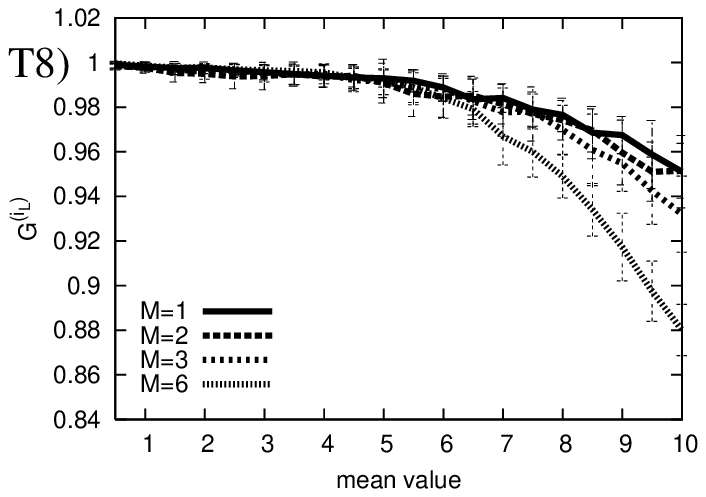}\\
\includegraphics[width=0.22\textwidth]{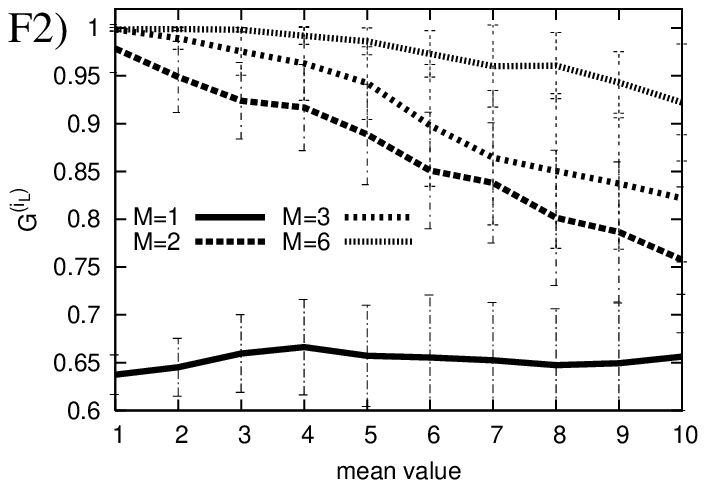}&
\includegraphics[width=0.22\textwidth]{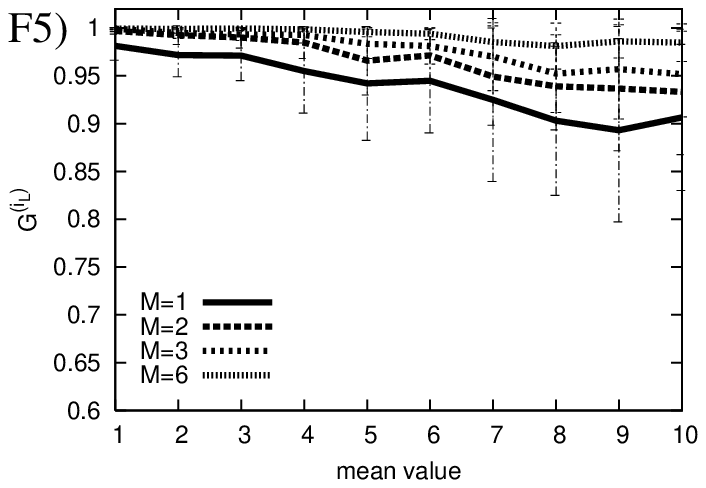}&
\includegraphics[width=0.22\textwidth]{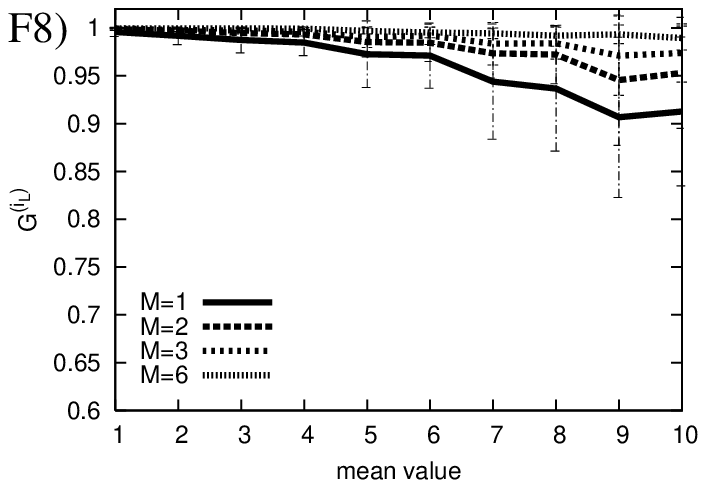}\\
\includegraphics[width=0.22\textwidth]{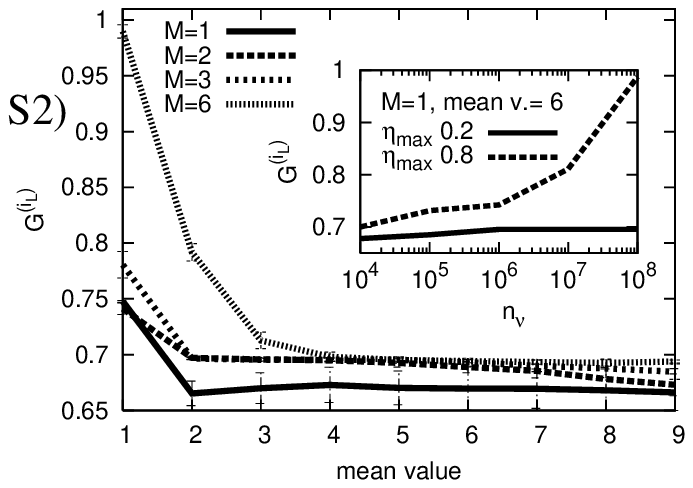}&
\includegraphics[width=0.22\textwidth]{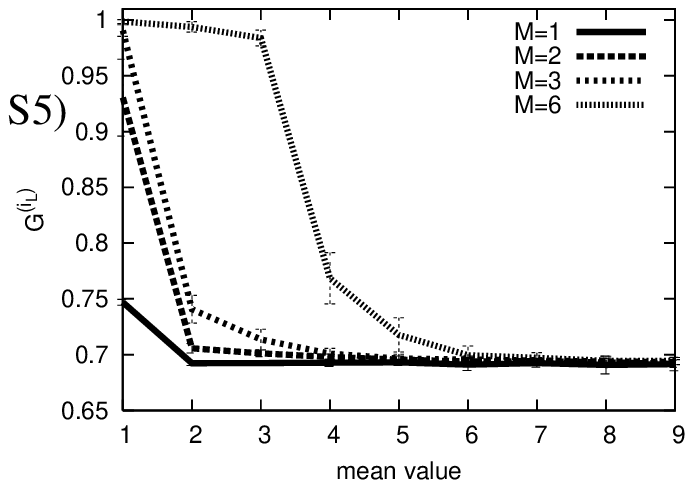}&
\includegraphics[width=0.22\textwidth]{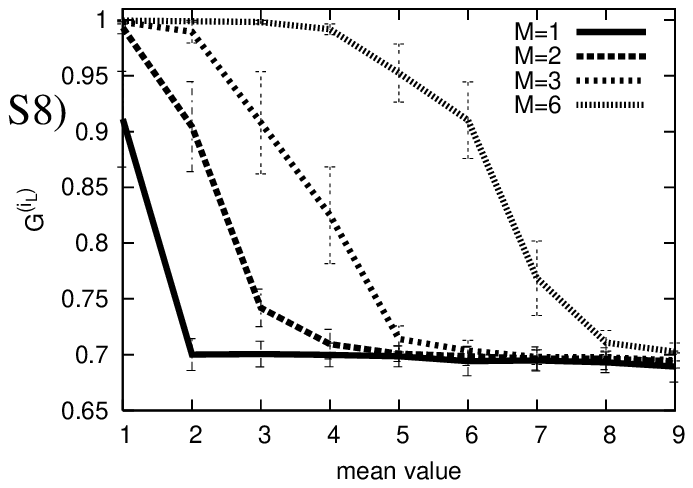}
\end{tabular}
\caption{Fidelity parameter at the last iteration, $G^{(i_{L})}$,
as a function of $\langle a^\dag a\rangle$ for $4$ different 
statistics of $\{\varrho_n\}$: coherent, panels labeled C);
thermal, label T); Fock state, label F); superposition of 2 
Fock states, $ \frac{1}{\sqrt{2}}(|\langle a^\dag a\rangle-
1\rangle + |\langle a^\dag a\rangle+1\rangle)$, label S).
The reconstructions have been performed using 
$\eta_{max}=0.2, 0.5$ and $0.8$,
panels labeled with 2), 5) and 8) respectively;
$n_\nu=10^4$ number of runs at each for every $\eta_\nu$.
According to the rule given in the text, the last iteration performed
is $\#(i_{L})=n_\nu$ except for coherent state with $M=1$ and
$\langle a^\dag a\rangle=1...4$ for which $\#(i_{L})=10^5$,
and for the superpositions state.
Inset of S2) show $G^{(i_{L})}$ for
$\frac{1}{\sqrt{2}}(|5\rangle + |7\rangle)$,
M=0, quantum efficiency 0.2 and 0.8 in function of $n_\nu$.
Every point of the curves is the mean of 40 simulations, the error bars
represent the standard deviations.
The other parameters are the same as in fig.\ref{fig:pois1e6}.}
   \label{fig:tanti}
\end{figure}
\end{widetext}
\par
Let us now illustrate our results about the accuracy of the
reconstructions. As mentioned above we performed simulated 
experiments, on semiclassical states like coherent and thermal 
ones as well as highly nonclassical states such Fock states, by
using three values for the maximum quantum efficiency 
$\eta_{\max}=0.2,0.5,0.8$.
Simulations have been performed by using $n_\nu=10^4$ data at each
value of the quantum efficiency.
This value of $n_\nu$ is relatively small and well within the realm
of quantum optical experiment.
In a real experiment performed under the same conditions,
$n_\nu$ of this order would allow on-line reconstruction of
the photon statistics.
Simulations performed using different values of the parameters
lead to similar conclusion.
\par
In Fig. \ref{fig:tanti} we show the fidelity of reconstruction
$G^{(i)}$ at the last iterations, $G^{(i_{L})}$,
for different input signals and different resolution thresholds 
as a function of the average number of photons of the signal.
As it is apparent from the plots, for low quantum efficiency and 
semiclassical states on/off detectors ($M=1$) are enough to achieve
a good reconstruction. On the other hand, the accuracy for 
nonclassical states largely improves for $M \geq 2$.
For higher values of the quantum efficiency (the middle and the 
right plots) on/off detectors becomes sufficient for a good
reconstruction also for nonclassical states having single-peaked distributions.
Notice that a higher resolution, however with a low value of 
$\eta_{max}$, does not guarantee a good reconstruction.
Also notice that the results reported in Fig. \ref{fig:tanti} 
has been obtained by stop the algorithm according to
the prescriptions mentioned above. We therefore do not expect 
that the fidelity is optimal for {\em any} state. 
Indeed the fidelity $G^{(i_{L})}$ in panel C5) and C8) of 
Fig. \ref{fig:tanti} slightly decreases for high number of
photons, though remaining close to unit value.
The decreases in panels T2), T5) and T8) of Fig. \ref{fig:tanti}
for greater mean values is due to the choice of a small 
dimension of the truncated Hilbert space:
by increasing N the trend disappear.
The low values of $G^{(i_{L})}$ for the superposition states are
due to the noise in $f^m_\nu$, also for the
high quantum efficiencies. Indeed, by increasing $n_\nu$ the 
reconstructions greatly improves; besides, for high quantum 
efficiency the detectors perform good reconstructions also for
$M=0$, see the inset of panel S2) in Fig. \ref{fig:tanti}.
The error bars in Fig. \ref{fig:tanti} are standard deviations
of $G^{(i_{L})}$ as calculated from 40 different Monte Carlo 
runs. They may appear large, but this is due to the high value
of $G^{(i_{L})}$, which in turn determines the scale of the plot. 
Look at panels S)' for comparison.
\section{Experimental data}\label{s:exp}
In order to confirm the Monte Carlo results for single-peaked
distributions we have performed the reconstruction of the photon
statistics of a coherent signal obtained from a Nd:YLF laser.
The experimental data have been recorded with a hybrid photo
detector, Hamamatsu H8236-40, placed on the second harmonics
(523.5 nm) of a cw mode-locked Nd:YLF laser regeneratively
amplified at a repetition rate of 5 kHz (High Q Laser).
\par
The frequencies $f_\nu^m$, until $m=3$, have been extrapolated 
from the response of the detector with the following procedure. 
At first a Gaussian best-fit of the response peaks at 0,..,4
photoelectrons has been performed \cite{burle}, then the
the frequencies has been obtained by choosing three thresholds, 
whose optimal choice turns out to be the mean point between
photoelectron peaks\cite{mariatagli}.
After the frequencies $f_\nu^m$ have been obtained we used 
the algorithm of the previous Section.
The results of the reconstruction are shown in Fig. 
\ref{fig:sperim}, panel a).
The reconstructed distribution at the last iteration
$\varrho^{(i_L)}_n$ have been then compared with a Poissonian best-fit.
As it is apparent from Fig. \ref{fig:sperim}, the agreement between
$\varrho^{(i_L)}_n$ and the best-fit is very good.
In addition, we notice that the shape of $G^{(i)}$ 
and $\varepsilon^{(i)}$ are very similar to those coming from 
simulation and that the $\varrho^{(i)}_n$ obtained by the experimental 
data with M=1 is close to the one obtained with M=2 and M=3,
as predicted by simulations.
\begin{widetext} $ $ 
\begin{figure}[h!]
\includegraphics[width=0.25\textwidth]{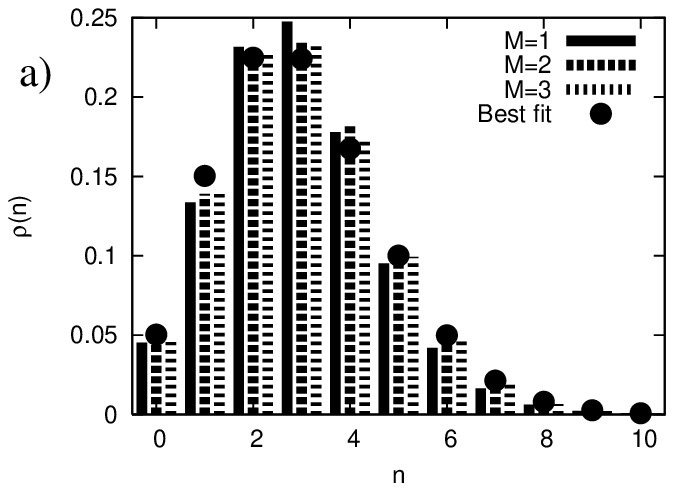}
\includegraphics[width=0.25\textwidth]{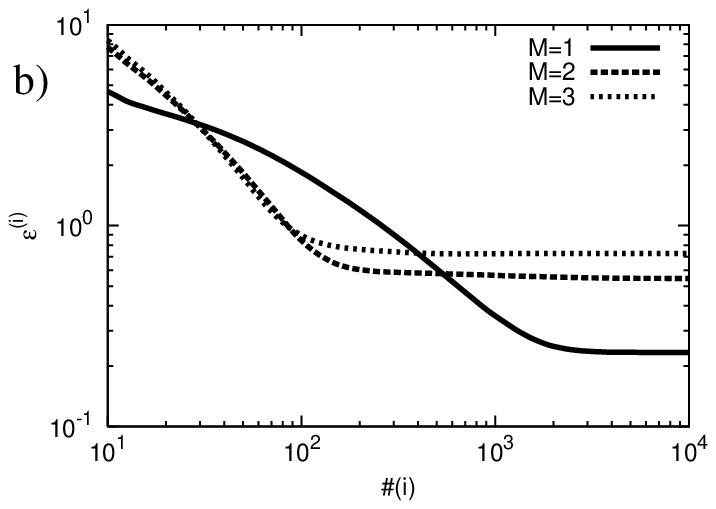}
\includegraphics[width=0.25\textwidth]{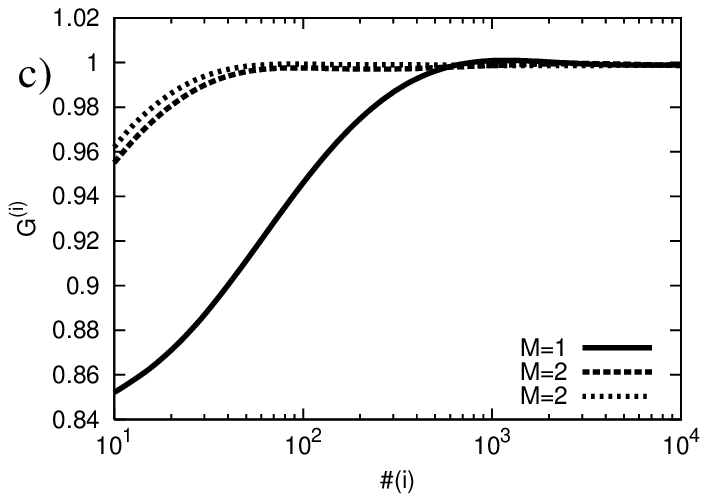}
\caption{Experimental reconstruction of the photon number distribution of 
a coherent state. a): reconstructed $\rho_n$ and coherent best fit,
$\langle a^\dag a\rangle=2.98$; b): total absolute error $\varepsilon^{(i)}$; 
c): fidelity $G^{(i)}$ calculated \textit{a posteriori}.
The reconstruction is performed with a hybrid photo detector operated 
by taking $M=1$ (on/off detector), 
$M=2$ and $M=3$; $K=100$ different quantum efficiencies 
$\eta=\eta_\nu$ distributed in $[0,\eta_{max}]$, 
$\eta_{max}=0.4$; the Hilbert space is truncated at $N=30$; $n_\nu=10^4$ 
number of runs have been performed for each $\eta$; The algorithm 
is stopped at iteration $\#(i_{L})=n_\nu$.}
\label{fig:sperim}
\end{figure}
\end{widetext}
\section{Conclusion}\label{s:out}
We have compared the reconstruction of the photon statistics 
as obtained applying ML algorithm to data coming from detectors 
with different counting capability and different quantum 
efficiencies. We found that using ML methods, detectors with 
high quantum efficiency does not need to
have high counting capability, since on/off detection assisted
by ML methods already provides good state reconstructions.  On
the other hand, a small quantum efficiency makes the counting
capability a crucial parameter.  Overall, our results indicate
that development of future photodetectors may be focused on
increasing the quantum efficiency rather than the counting capability. 
\section{Acknowledgments}\label{Acknow}
The authors are grateful to Alessandra Andreoni for encouragement 
and support. They also thank Maria Bondani, Fabio Ferri, Zdenek 
Hradil and Jarda \v{R}eh\'{a}\v{c}ek for useful discussions. This 
work has been partially supported by MIUR through the projects 
PRIN-2005024254-002 and FIRB RBAU014CLC-002.


\begin{thebibliography}{99}
\bibitem{pst}E. Knill, R. Laflamme and G.J. Milburn, Nature (London)
M. Dakna, J. Clausen, L. Knoll, D.-G. Welsch, 
Phys. Rev. A {\bf 59}, 1658 (1999); 
M. G. A. Paris, Phys. Rev. A. {\bf 62}, 033813 (2000);
M. G. A. Paris, M. Cola, R. Bonifacio, 
Phys. Rev. A {\bf 67}, 042104 (2003); 
J. Laurat et al., Phys. Rev. Lett. {\bf 91}, 213601 (2003);
S. A. Babichev, B. Brezger, A. I. Lvovsky, Phys. Rev. Lett. {\bf 92}, 
047903 (2004).
\bibitem{post}E. Knill, R. Laflamme and G.J. Milburn, Nature (London)
{\bf 409}, 46 (2001)
\bibitem{rivelacontaS}  D. Rosenberg, A.E. Lita, A.J. Miller, S.W. Nam,
Phys. Rev. A {\bf 71}, 61803 (2005).
\bibitem{rivelacontaH}
R. DeSalvo, Nucl. Instr. and Meth. A 387 (1997) 92-96.
\bibitem{burle} G. Zambra, M. Bondani, A. S. Spinelli, A. Andreoni,
Rev. Sci. Instrum. {\bf 75}, 2762 (2004).
\bibitem{NIST} E. Hergert, Single Photon Detector Workshop, Gaithersburg, NIST (2003).
\bibitem{xxx} J. Kim, S. Takeuchi, Y. Yamamoto, and H.H. Hogue, Appl.
Phys. Lett. {\bf 74}, 902 (1999); A. Peacock, P. Verhoeve, N. Rando,
A. van Dordrecht, B. G. Taylor, C. Erd, M. A. C. Perryman, R. Venn,
J. Howlett, D. J. Goldie, J. Lumley, and M. Wallis, Nature {\bf
381}, 135 (1996).
\bibitem{serg} G. Di Giuseppe, A. V. Sergienko, B. E. A. Saleh, and
M. C. Teich in {\em Quantum Information and Computation}, E. Donkor,
A. R. Pirich, and H. E. Brandt Eds., Proceedings of the SPIE {\bf
5105}, 39 (2003).
\bibitem{mun} M. Munroe, D. Boggavarapu, M. E. Anderson, and M. G. Raymer,
Phys. Rev. A {\bf 52}, R924 (1995); Y. Zhang, K. Kasai, and M.
Watanabe, Opt. Lett. {\bf 27}, 1244 (2002).
\bibitem{deconvolvphoton} T. Kiss, U. Herzog, U. Leonhardt,
Phys. Rev. A {\bf 52}, 2433 (1995); G. M. D'Ariano, C. Macchiavello, Phys.
Rev. A 57, 3131 (1998); T. Kiss, U. Herzog, and U. Leonhardt, Phys. Rev.
A 57, 3134 (1998).
\bibitem{mogy} D. Mogilevtsev, Opt. Comm {\bf 156}, 307 (1998); Acta Phys.
Slov. {\bf 49}, 743 (1999).
\bibitem{pcount}  A. R. Rossi, S. Olivares, and M. G. A. Paris,
Phys. Rev. A {\bf 70}, 055801 (2004).
\bibitem{cvp} G. Zambra et al, Phys. Rev. Lett., 
{\bf 95}, 063602 (2005).
\bibitem{olom} J.  $\check{\rm R}$eh$\acute{\rm a}\check{\rm c}$ek, Z.
Hradil, O. Haderka, J.  Pe$\check{\rm r}$ina, Jr., and M.  Hamar, Phys.
Rev. A {\bf 67}, 061801(R) (2003); O. Haderka, M. Hamar, J. Pe$\check{\rm
r}$ina, Eur. Phys. Journ.  D {\bf 28}, (2004).
\bibitem{phc} H. Paul at al., Phys. Rev. Lett. {\bf 76}, 24642467 (1996)
\bibitem{EM:alg:1} A.P. Dempster, N.M. Laird, D.B. Rubin, J. R. Statist. Soc. B
{\bf 39}, 1 (1977); Y. Vardi and D. Lee, J. R. Statist. Soc. B
{\bf 55}, 569 (1993).
\bibitem{EM:alg:2} R. A. Boyles, J. R. Statist. Soc. B {\bf 45},
47 (1983); C. F. J. Wu, The Annals of Statistics {\bf 11}, 95
(1983).
\bibitem{kon} K. Banaszek, Acta Phys. Slov. {\bf 48}, 185 (1998); 
Phys. Rev. A {\bf 57}, 5013 (1998).
\bibitem{mariatagli} M. Bondani, \emph{Self consistent characterization
of light statistics}, manuscript in preparation.
\end{thebibliography}
\end{document}